\documentstyle[12pt,a41,epsf]{article}

\newcommand{\be}{\begin{equation}}
\newcommand{\ee}{\end{equation}}
\newcommand{\ba}{\begin{eqnarray}}
\newcommand{\ea}{\end{eqnarray}}
\newcommand{\bb}{\begin{thebibliography}}
\newcommand{\eb}{\end{thebibliography}}
\newcommand{\ci}[1]{\cite{#1}}
\newcommand{\bi}[1]{\bibitem{#1}}
\newcommand{\lab}[1]{\label{#1}}

\begin{document}

\begin{center}
{\LARGE\bf Spin  Effects  In Diffractive Processes  at HERA}

\vspace{1cm}
{S.V.Goloskokov $^a$}

\vspace*{1cm}
{\it $^a$ Bogoliubov Laboratory of Theoretical  Physics,
  Joint Institute for Nuclear Research,\\Dubna 141980, Moscow region, Russia 
}\\

\vspace*{2cm}

\end{center}

\begin{abstract}
It is shown that the  $A_{ll}$
asymmetry in diffractive $Q \bar Q$ leptoproduction 
 is not small at HERA energies and dependent on the 
structure of the pomeron couplings and on the masses of  produced
quarks.  The connection of this asymmetry with the non-forward gluon
distribution in the proton is discussed.
\end{abstract}

\vspace{1mm}
\noindent
Study of diffractive processes at HERA provides an excellent
possibility to investigate the nature of the pomeron \ci{h1_zeus,lev}. 
Investigation of the  vector meson production becomes popular now.  
On the one hand, these reactions give 
 information on the pomerom structure, and on the other
hand, they can be used to analyze the gluon distribution $ G(x)$
 at small $x$ \ci{rys,diehl}. 
 The longitudinal double-spin asymmetry
in these processes might be proportional to $[\Delta G/G]$. This
have initiated  proposals to test the polarized gluon
distribution $\Delta G$ from the $A_{LL}$ asymmetry in $J/\Psi$
production  \ci{anselm}.

However, it has been shown in \ci{gola_ll} that the double-spin asymmetry
in diffractive processes is proportional to the fraction  of the initial
proton momentum $x_p$ carried off by the pomeron. 
For the case of diffractive  $J/\Psi$
production $x_p$ is 
fixed  by the reaction kinematics: $x_p \sim (m_J^2+Q^2+|t|)/(s y)$.
As a result, the relevant $A_{ll}$ asymmetry  should be very small at HERA 
energies.

In this report, we shall study the $A_{ll}$
asymmetry for diffractive $Q \bar Q$ leptoproduction
of  light and heavy quarks.
In these processes, the $A_{ll}$  asymmetry is not small at high energies  
\ci{gola_ll}  because $x_p$ is not fixed in this case.
It will be shown that this asymmetry is sensitive to the spin
structure of the pomeron couplings and to the mass of the produced
quarks.

The pomeron  is a vacuum $t$-channel exchange that contributes to high-energy
diffractive reactions.  The hadron-hadron scattering amplitude determined by 
the pomeron exchange can be written in the form
$$
T(s,t)^{A,B}=I\hspace{-1.6mm}P(s,t) V_{AI\hspace{-1.1mm}P}^{\mu} \otimes
V^{BI\hspace{-1.1mm}P}_{\mu},    $$
where $I\hspace{-1.6mm}P$ is a "bare" pomeron contribution, and
$V_{AI\hspace{-1.1mm}P}$ and $V^{BI\hspace{-1.1mm}P}$ are the
pomeron vertices for particles $A$ and $B$, respectively.

When the gluons from the pomeron couple to a single quark in the hadron, 
 a simple matrix structure of pomeron vertex
$
V^{\mu}_{hh I\hspace{-1.1mm}P} =\beta_{hh I\hspace{-1.1mm}P}
\gamma^{\mu}$ appears.
 This standard coupling leads to spin-flip effects decreasing
 with energy like $1/s$.

The large-distance loop contributions  complicate the spin
structure of the pomeron coupling. These effects are determined by
the hadron wave function for the pomeron-hadron couplings or by the
gluon-loop corrections for the quark-pomeron case.

The spin structure of the quark-pomeron coupling
$V_{qqI\hspace{-1.1mm}P}^{\mu}$ has been studied in \ci{golpl}. It has
been shown that in addition to the standard pomeron vertex
  the gluon-loop contributions are important which
have provided the following form
of the quark-pomeron vertex:
\begin{equation}
V_{qqI\hspace{-1.1mm}P}^{\mu}(k,r)=\gamma^{\mu} u_0+2 M_Q k^{\mu}
u_1+
2 k^{\mu}
/ \hspace{-2.3mm} k u_2 + i u_3 \epsilon^{\mu\alpha\beta\rho}
k_\alpha r_\beta \gamma_\rho \gamma_5+i M_Q u_4
\sigma^{\mu\alpha} r_\alpha,    \label{ver}
\end{equation}
where $k$ is the quark momentum, $r$ is the momentum transfer and
$M_Q$
is the quark mass.  The functions
$u_1(r) - u_4(r)$ have been calculated perturbatively and 
can reach $30 \div 40 \%$ of the $u_0(r)$  term
 for $|r^2| \simeq {\rm few}~ GeV^2$ \ci{golpl}.
They  lead to spin-flip effects in the quark-pomeron coupling which do not
decrease with energy growth.

Let us  analyze the effects of the quark--pomeron
coupling  in the polarized diffractive $e+p \to e'+p'+Q \bar Q$
reaction and estimate  the longitudinal
double--spin $A_{ll}$ asymmetry  for light and heave quark
production at HERA energies.  The $A_{ll}$ asymmetry  is
determined by the relation
$A_{ll}=
(\sigma(^{\rightarrow} _{\Leftarrow})-\sigma(^{\rightarrow}
_{\Rightarrow}))/
(\sigma(^{\rightarrow} _{\Rightarrow})+\sigma(^{\rightarrow}
_{\Leftarrow}))$
where $\sigma(^{\rightarrow} _{\Rightarrow})$ and
$\sigma(^{\rightarrow}
_{\Leftarrow})$ are the cross sections with parallel and antiparallel
longitudinal polarization of lepton and proton.

The  trace over the quark loop for the difference of the polarized cross
sections for  the standard pomeron vertex looks like
\be
 A^s(\beta,k_\perp^2,t)=
   16 (2 (1-\beta) k_\perp^2 - |t| \beta - 2 M_Q^2 (1+\beta))
|t|.
  \lab{ad}
 \ee
Similar forms can be written for the spin--average cross
sections.
In  all the cases  the strong dependence of the cross sections on
the mass of the produced quarks and $\beta \simeq Q^2/(Q^2+M_x^2)$ 
has been observed. Relevant functions for the spin-dependent vertex and the 
explicit forms for the cross sections can be found in \ci{gola_ll,g_all}.

\begin{minipage}{8cm}
\epsfxsize=7.9cm
\epsfbox{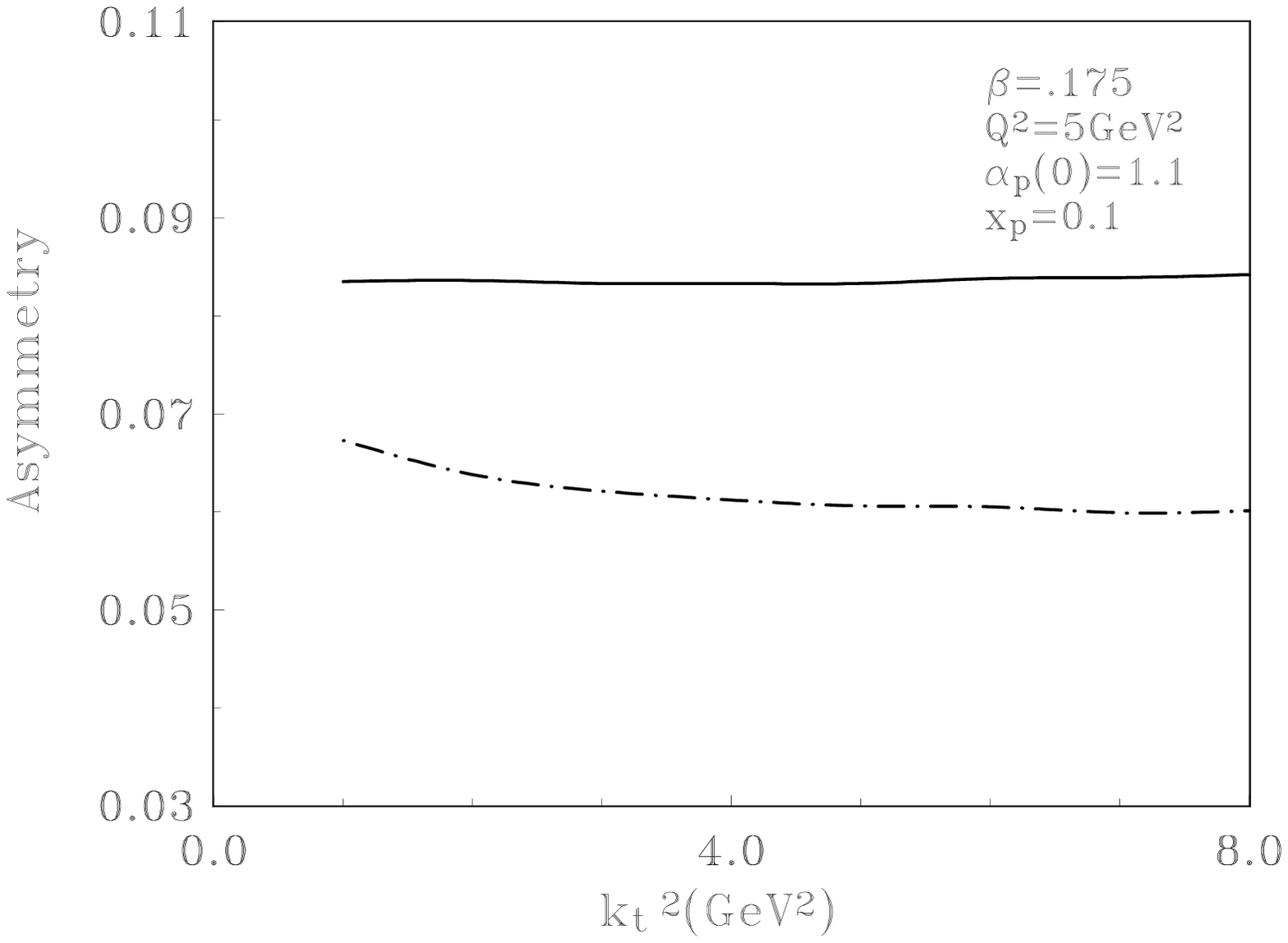}
\end{minipage}
\begin{minipage}{0.5cm}
\end{minipage}
\begin{minipage}{8cm}
\epsfxsize=7.9cm
\epsfbox{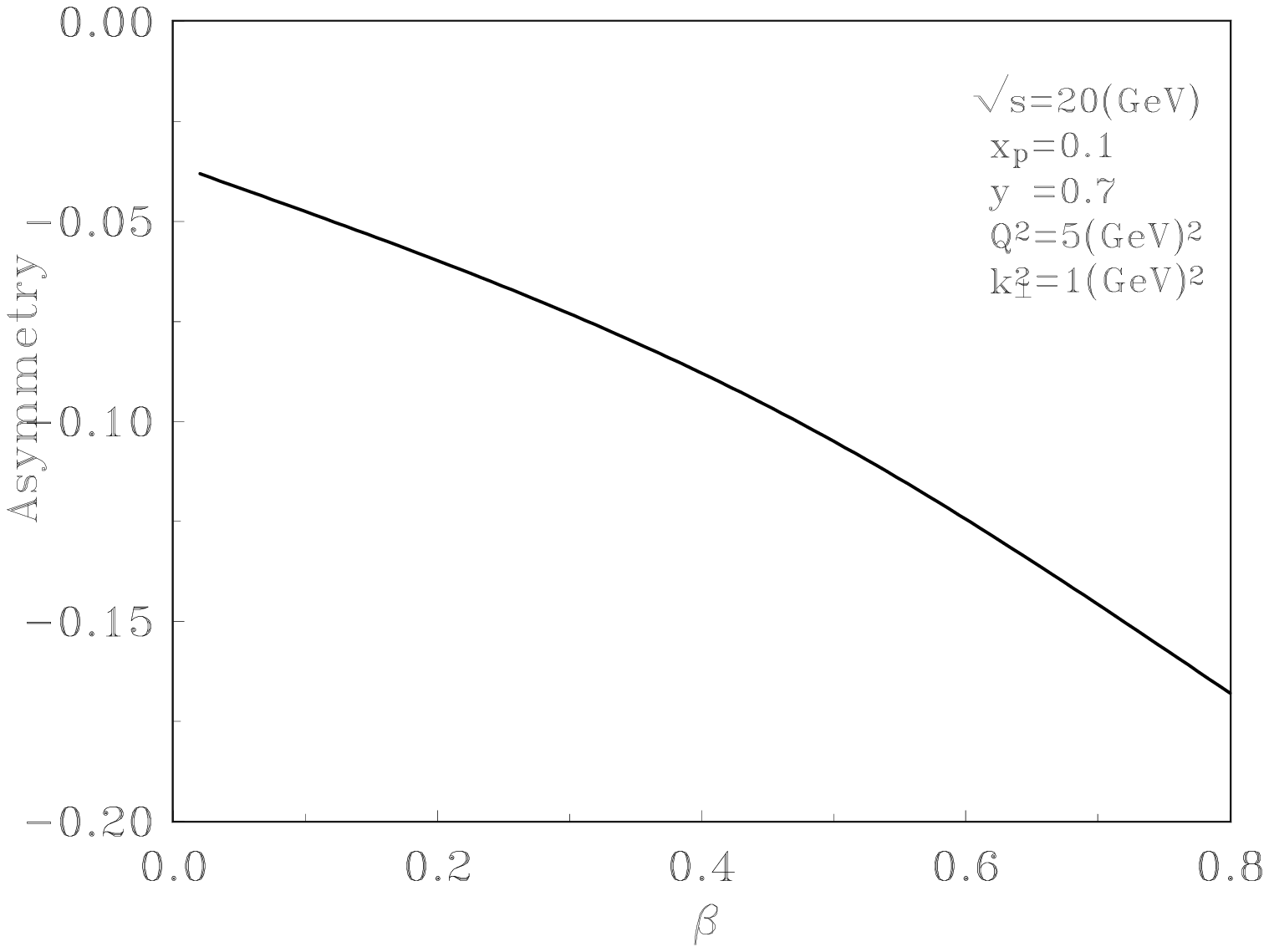}
\end{minipage}

\begin{minipage}{8cm}
Fig.1~ $A_{ll}$ asymmetry of light quark
production.
Solid line -for the standard vertex;
dot-dashed line -for the spin-dependent  vertex.
\end{minipage}
\begin{minipage}{0.5cm}
\end{minipage}
\begin{minipage}{8cm}
Fig.2~$\beta$-- dependence of $A_{ll}$ asymmetry for diffractive
open charm production for the standard  quark-pomeron vertex.
\end{minipage}
\bigskip
 
We calculate the cross section
 integrated over the pomeron momentum transfer because it is usually
difficult to
detect the recoil proton in diffractive experiments.
The asymmetry of the diffractive light $Q \bar Q$ production is
shown in Fig. 1.
The asymmetry for the standard quark--pomeron vertex is very
simple in form
\be
A_{ll}=\frac{y x_p (2-y)}{2-2y+y^2}.
\ee
There is no any $k_\perp$ and $\beta$ dependence here. For the
spin--dependent pomeron coupling the
 $A_{ll}$ asymmetry is smaller than for the standard
 pomeron vertex and depends on $k^2_{\perp}$.
Thus, it is sensitive to the spin structure of the
pomeron coupling.

The predicted $A_{ll}$ asymmetry for the diffractive open charm ($c \bar c$)
production  is large and negative (Fig.2). 
It can be seen that to obtain  information about the pomeron--coupling 
structure,
the relevant asymmetries should be measured with  accuracy less than
1\%.  The estimations for the cross sections \ci{gola_ll,g_all} show 
that such statistical errors  can be obtained
at the integrated luminosity about $200 (pb)^{-1}$.

The large magnitude of asymmetry in diffractive $Q \bar Q$ production is
caused by  not small $x_p \sim 0.1$ which is typical of these reactions.
This means that in contrast with the vector meson production,
 where the smallness 
of $x_p$ permits one to connect the two-gluon structure of the 
pomeron with $G(x)$, the
diffractive $Q \bar Q$ production might give information on the 
non-forward gluon distributions in the proton.

We have found that the spin structure of the pomeron
coupling should modify the
spin average and  spin--dependent cross section in 
diffractive processes.
Not small values for the $A_{ll}$ asymmetry in the diffractive $Q
\bar Q$ production have been predicted. 
The asymmetry is free from the normalization
factors and is sensitive to the dynamics of pomeron interaction.
 Thus, the $A_{ll}$ asymmetry in diffractive $ Q \bar Q$
leptoproduction is convenient to test the pomeron coupling
structure. Moreover, it can be connected with the the non-forward spin-dependent 
gluon distributions.

Thus, we can conclude that the pomeron coupling structure
can be studied in  diffractive processes.
Note that the spin--structure of the pomeron vertex is
determined by
the large--distance contributions. 
So, the important test of the spin structure of QCD at
large distances can be carried out by studying diffractive
reactions in future polarized experiments     at HERA.

This work was supported in part by the Heisenberg-Landau Grant.

\end{document}